\documentclass[a4paper]{article}

\usepackage[english]{babel}
\usepackage[utf8x]{inputenc}
\usepackage[T1]{fontenc}
\usepackage{multirow}
\usepackage[a4paper,top=2.5cm,bottom=2.5cm,left=2.3cm,right=2.3cm,marginparwidth=1.75cm]{geometry}

\usepackage{amsmath}
\usepackage{graphicx}
\usepackage[colorinlistoftodos]{todonotes}
\usepackage[colorlinks=true, allcolors=blue]{hyperref}

\title{Cosmic Visions Dark Energy: Small Projects Portfolio}
\author{Cosmic Visions Dark Energy Panel: Kyle Dawson, Josh Frieman, Katrin Heitmann, \\ Bhuvnesh Jain, Steve Kahn, Rachel Mandelbaum, Saul Perlmutter, An\v{z}e Slosar }

\begin{document}
\maketitle

\thispagestyle{empty}


\section*{Executive Summary}

The 2014 P5 Report identified understanding cosmic acceleration as one of the key science drivers for high-energy physics in the coming decade. With the Large Synoptic Survey Telescope (LSST) and the Dark Energy Spectroscopic Instrument (DESI) beginning operations soon, we are entering an exciting phase during which we expect an order of magnitude improvement in constraints on dark energy and the physics of the accelerating Universe.
This is a key moment for a matching Small Projects portfolio that can (1) greatly enhance the science reach of these flagship projects, (2) have immediate scientific impact, and (3) lay the groundwork for the next stages of the Cosmic Frontier Dark Energy program. In this White Paper, we outline a balanced portfolio that can accomplish these goals through a combination of observational, experimental, and theory and simulation efforts.  The portfolio includes:

\begin{itemize}
\item {\bf Observations that leverage existing facilities} to expand the dark energy science reach of LSST and DESI.  Complementary data from spectroscopic, space-based, and multi-wavelength facilities will provide insights into LSST and DESI data, thereby reducing both statistical and systematic uncertainties compared to their baseline programs. This program of observations will need to be complemented by support for members of the Cosmic Frontier community to integrate analysis of these other samples with the analysis of LSST and DESI to enhance the constraints on dark energy, inflation, and neutrino masses;
\item {\bf Experimental R\&D} to address the key technical challenges the community faces when looking forward to the next generation of dark energy experiments; 
\item{\bf Theory and simulation development} that will allow us to explore and extract cosmological information from small scales and develop novel probes from dark energy surveys. Improved modeling of survey observables on moderately nonlinear scales will pay direct dividends in providing stronger cosmological constraints. 
Theoretical exploration of the highly nonlinear regime, and more sophisticated modeling of astrophysical effects, unlocks the small-scale cosmological information content of LSST and DESI.  Models in these regimes can potentially yield constraining power that vastly exceeds the standard projections for constraints on dark energy, inflation, and neutrino masses.
The theoretical exploration and substantial program of simulations of structure formation that this will require will also enable us to explore the unknown in terms of new interactions, fields, and physical principles guiding cosmic expansion.
\end{itemize}

The Small Projects portfolio described below follows the general themes laid out in two earlier Cosmic Visions White Papers \cite{2016arXiv160407626D,2016arXiv160407821D} and grew out of two community workshops that followed them, the KICP Future Cosmic Surveys Workshop in Sept. 2016\footnote{http://kicp-workshops.uchicago.edu/FutureSurveys/} and the LBNL Cosmic Visions Workshop: Dark Energy in Nov. 2017\footnote{http://cvde2017.lbl.gov/}. During these workshops, multiple small projects were identified as part of the three overarching themes outlined above.  As shown in Appendix~A, the required support for most of these projects is in the \$1M-\$3M regime, leading to a modest overall investment in the \$10M range. The support of several, carefully-chosen, small-scale efforts will have a greater impact than supporting a single larger one, since they provide complementary paths for attacking the complex problem posed by dark energy. These projects can be brought online rapidly, on a timescale that best supports LSST and DESI. Now is therefore the time to build such a portfolio of research, R\&D, and observations in the Cosmic Frontier community, in order to enhance and extend the existing projects and lay the groundwork for the future.\footnote{The report is based on contributions from: Tom Abel, Zeeshan Ahmed, Greg Aldering, Sahar Allam, Lori Allen, David 
Alonso, Marcelo Alvarez, Adam J Anderson, Jim Annis, Reza Ansari,
Rafael Arcos-Olalla, Charles Baltay, Darcy Barron, Keith Bechtol,
Andrew Benson, Jonathan Blazek, Lindsey Bleem, Sebastian Bocquet, Adam
Bolton, Kyle Boone, Andrew Bradshaw, Elizabeth Buckley-Geer, Philip
Bull, Kevin Bundy, Robert Cahn, Robert Caldwell, Peter Capak, Chris
Carilli, Emanuele Castorina, Tzu-Ching Chang, Swapan Chattopadhyay,
Xuelei Chen, Asantha Cooray, Neal Dalal, Marc Davis, Kyle Dawson, Will
Dawson, Darren DePoy, Roland de Putter, Marcellinus Demarteau, Tom
Diehl, Samantha Dixon, Scott Dodelson, Olivier Doré, Alex
Drlica-Wagner, Juan Estrada, Giulio Fabbian, Simone Ferraro, David
Finley, Brenna Flaugher, Simon Foreman, Wendy Freedman, Josh Frieman,
Josef Frisch, Enrique Gaztanaga, Mandeep Gill, Fred Gilman, Danny
Goldstein, Daniel Green, Ravi Gupta, Gaston R Gutierrez, Salman Habib,
ChangHoon Hahn, Patrick Hall, Andrew Hearin, Katrin Heitmann,
Jacqueline Hewitt, Christopher Hirata, Renee Hlozek, Shirley Ho, Steve
Holland, Daniel Holz, Klaus Honscheid, Dragan Huterer, Bhuvnesh Jain,
Daniel Jacobs, Stephanie Juneau, Elise Jennings, Garrett K. Keating,
Robert Kehoe, Stephen Kent, Gourav Khullar, Alex Kim, David Kirkby,
Eve Kovacs, Ely Kovetz, Elisabeth Krause, Richard Kriske, Richard
Kron, Steve Kuhlmann, James E Lasker, Alexie Leauthaud, Ricky Leavell,
Khee-Gan Lee, Boris Leistedt, Michael Levi, Ting Li, Adam A Lidz, Huan
Lin, Eric Linder, Jessica Lu, Zarija Lukic, Niall F MacCrann, Rachel
Mandelbaum, Alessandro Manzotti, Jennifer Marshall, Phil Marshall,
Paul Martini, Patrick McDonald, Peter Melchior, Joel Meyers, Kavilan
Moodley, Miguel F Morales, Pavel Motloch, Li Nan, Laura Newburgh,
Jeffrey Newman, Peder Norberg, Paul O'Connor, Nikhil Padmanabhan,
Samuel Passaglia, Jeff Peterson, Jonathan Pober, Jason Poh, Kara
Ponder, Anthony R. Pullen, Matt Radovan, Alexandra S Rahlin, Marco
Raveri, Paul Ricker, Constance Rockosi, Natalie Roe, Aaron Roodman,
Ashley Ross, Aditya Rotti, Khaled Said, David Schlegel, Marcel M
Schmittfull, Michael Schneider, Daniel Scolnic, Douglas Scott, Uros
Seljak, Hee-Jong Seo, Richard Shaw, Christopher Sheehy, Nora Shipp,
David Silva, Sukhdeep Singh, Zachary Slepian, Anze Slosar, Marcelle
Soares-Santos, Tony Spadafora, Paul Stankus, Albert Stebbins, Michael
Strauss, Christopher Stubbs, Meng Su, Amy Tang, Tommaso Treu, Mark
Trodden, Douglas Tucker, Tony Tyson, Amol Upadhye, Anja van der
Linden, Mohammadjavad Vakili, Ricardo Vilalta, Francisco
Villaescusa-Navarro, Benjamin Wandelt, Xin Wang, Risa Wechsler, Martin White,
Michael Wilson, Friedwardt Winterberg, Kimmy Wu, Christophe Yeche, Jun
Zhang, Yuanyuan Zhang, Idit Zehavi, Zheng Zheng
}




\newpage
\setcounter{page}{1}
\section{Motivation}
\label{sec:mot}

The discovery of late-time cosmic acceleration has profound ramifications for fundamental physics, as recognized by the 2011 Nobel Prize in Physics. Currently, the three main contenders to explain this phenomenon are
(i) a cosmological constant $\Lambda$, which can alternately be described as a dominant stress-energy component with a constant equation of state parameter, $w=-1$, (ii) a dynamical dark energy component with a different and typically time-varying equation of state parameter, $w(a)\neq -1$, and (iii) a modification of Einstein's theory of General Relativity on cosmological scales. Lacking a compelling theoretical model for cosmic acceleration at the current time, our field is driven by observations that aim in part to test the cosmological constant plus cold dark matter model ($\Lambda$CDM) and to 
find hints of new physics beyond it. 

Our primary means of gaining information about the underlying mechanism of cosmic acceleration are the cosmic expansion history and the evolution of cosmic structure. The propagation and deflection of light provides a complementary means to test gravitation on cosmological scales. The dark energy community has developed a comprehensive program of imaging and spectroscopic surveys designed to measure expansion history, structure growth, and light deflection through multiple observables such as gravitational lensing, galaxy clusters, the large-scale distribution of galaxies, and type Ia supernovae. The program follows the staged approach described by the Dark Energy Task Force (DETF,~\cite{2006astro.ph..9591A}), with sequential stages characterized by an increasing figure of merit (FoM) for constraining the dark energy equation of state parameter and its time evolution. Currently, we are in the era of Stage III experiments; DOE supports the two most prominent programs currently operating, DES and eBOSS, which are forerunners of the Stage IV projects LSST and DESI.  

The Dark Energy Survey (DES) is utilizing the Dark Energy Camera \cite{flaugher15a} on the 4-meter Blanco telescope in Chile to conduct a 5-year, multi-band imaging survey unprecedented in its combination of depth and breadth. DES is designed to employ all the dark energy observables mentioned above, along with cross-correlation with the Cosmic Microwave Background (CMB), and has reported cosmological results from its first year (Y1) of data using weak lensing and galaxy clustering measurements~\cite{2017arXiv170801530D,2017arXiv171206209T}. The Baryon Oscillation Spectroscopic Survey (BOSS) \cite{dawson13a} concluded in 2014; its successor, eBOSS \cite{dawson16a}, continues to use spectroscopy of galaxies and quasars to trace the underlying density field.  BOSS and eBOSS probe the largest volume to date of any cosmological redshift survey and provide constraints on both the distance scale through Baryon Acoustic Oscillations (BAO) and on the growth of structure through redshift-space distortions (RSD).  This combined spectroscopic survey provides measurements of BAO over a redshift range $0.2<z<3.5$, providing a view of cosmic expansion over two-thirds of the history of the Universe.

The early results from the Stage III experiments show broad consistency with the $\Lambda$CDM paradigm so far. Combining the DES Y1 results with Planck CMB measurements, SN Ia data, and BAO measurements including BOSS yields $w=-1.00^{+0.04}_{-0.05}$ in the context of the $w$CDM model with constant dark energy equation of state parameter~\cite{2017arXiv170801530D}. This result is robust to assumptions about spatial curvature: with curvature as a free parameter, the BOSS distance scale measurements combined with Planck yield $w = -1.01 \pm 0.06$ \cite{alam16a}. Some of the current experiments do show hints of tension in the inference of cosmological parameters from different observables, but the statistical significance is in most cases marginal. For example, BAO measurements from the final Lyman-$\alpha$ forest sample from BOSS (mean redshift $z=2.4$) \cite{bautista17a, masdesbourboux17a} yield constraints on the distance scale that lie $2.3 \sigma$ from the predictions of the flat $\Lambda$CDM model best fit by  the Planck data \cite{planck16a}.  
In addition, combining Planck, BOSS, and SN Ia data~\cite{alam16a}, or combining DES Y1, BOSS, and Big Bang Nucleosynthesis results~\cite{2017arXiv171100403D} yields a value of the Hubble parameter, $H_0 = 67.3 \pm 1.0$ km/s/Mpc, in tension with `direct' distance-scale measurements that do not assume $\Lambda$CDM.  For example, measurements using Cepheid variables and nearby Type Ia supernovae imply a value $H_0 = 73.2 \pm 1.7$ km/s/Mpc \cite{2016ApJ...826...56R}, while time-delay measurements from several strongly lensed quasars yield $H_0=71.9^{+2.4}_{-3.0}$ km/s/Mpc~\cite{2017MNRAS.468.2590S,2017MNRAS.465.4914B}.  There have also been hints of tension between weak lensing surveys and the best-fitting models from Planck in their constraints in the $(\Omega_m, \sigma_8)$ plane, though the significance of this tension depends on the survey (\cite{2017MNRAS.471.4412K} report $3\sigma$ tension between Planck and KiDS, while the DES Y1 results \cite{2017arXiv170801530D} differ in the same direction from Planck but not to a statistically significant degree).  Both of the aforementioned tensions may signal the presence of new physics beyond $\Lambda$CDM, unknown systematic errors in one or more of the measurements, or a statistical fluctuation. More measurements with increasing precision and accuracy are required to better understand the origins of these discrepancies.  

The Stage III results highlight one of the key tasks for the dark energy community as we head toward Stage~IV and beyond -- to continue to develop robust analysis, observational, and theoretical programs to ensure that we can disentangle systematic effects from potential new physics.  LSST and DESI are currently being constructed as Stage~IV dark energy experiments. LSST will be a dedicated, large-aperture telescope that will obtain deep, multi-band images of nearly half the sky to great depth over the course of a decade-long observing campaign, with first light around 2020.  Like DES, LSST will constrain the nature of dark energy via weak and strong gravitational lensing, cluster counts, type Ia supernovae, and large-scale structure.  
Beginning in Fall 2019, DESI \cite{desi16a,desi16b} will carry out a spectroscopic redshift survey of galaxies and quasars that is an order of magnitude larger than that surveyed by the BOSS and eBOSS experiments.  The DESI design and science requirements were established to probe the origin of cosmic acceleration by using the BAO technique to measure the expansion history of the Universe.  
As imaging and spectroscopic surveys respectively, LSST and DESI are outstanding, complementary, stand-alone experiments. They can meet their high-level science requirements and goals without additional information or data.  However, the datasets are richer than the baseline metrics would lead one to believe.  

The drivers for our scientific program described in this report are aimed toward enhancing and extending the reach of the upcoming Stage IV experiments and toward laying critical groundwork for Stage V experiments. In particular, the efforts described are targeted at the following aims:

\begin{itemize}
\item The reduction of both statistical and systematic uncertainties for LSST and DESI beyond the current baseline to enhance cosmological constraints on dark energy, inflation, and neutrinos. Since the inception of the LSST and DESI concepts, the most important sources of systematic uncertainties have been better characterized, and we can now design a small program to help address these sources in a targeted way. Small-scale investments in calibration and efforts to lower the systematics floor due to known effects through specific, targeted observations could have a large payoff.  

\item The exploration of new probes and reduction of systematic uncertainties by cross-correlation between LSST and DESI and other multi-wavelength surveys.  In this case, our proposed program does not involve acquiring new data.  Rather, we propose to invest in an effort to develop methodologies to bring together data from DESI or LSST with external datasets for improved systematics control and/or better cosmological constraining power.  This can also include efforts to ensure complementarity in survey strategy between these and other surveys (such as WFIRST) in a way that would enhance the scientific gains from both, resulting in a `whole is greater than the sum of its parts' outcome. 

\item The exploration of small scales beyond the current DESI and LSST baselines by enabling a comprehensive modeling and simulation program. It has been suggested for some time that if the effects of gravitational collapse and baryonic matter can be encapsulated by a combination of empirical and theoretical modeling, then extending probes of large-scale structure well into the nonlinear regime can significantly improve dark energy constraints, even when self-calibrating a reasonably large number of ``nuisance parameters'' \cite{huterer_etal06,zentner_etal13}. Cosmological observations in the nonlinear regime also create many new opportunities to constrain models of modified gravity, and potentially falsify General Relativity, through  sensitive tests of the self-consistency of cosmic structure growth and expansion history~\cite{linder05,reid_etal14,zu_etal14,joyce_etal14}. 
We see no fundamental obstacle prohibiting the practical realization of the large science returns offered by probing cosmology in the nonlinear regime. The most significant impediment is probably people-hours: bringing these methods to scientific maturity prior to the arrival of Stage IV data will require a meaningful investment in personnel at the theory-simulation interface  beyond current levels.  

\item The investigation of novel probes. The small scale modeling effort above refers primarily to two-point correlations on nonlinear scales. However, there is a dazzling variety of astrophysical observations on small scales, and a number of these have been identified as offering sharp tests of fundamental physics beyond $w$CDM (i.e., beyond smooth dark energy models). Galaxy surveys, including DESI and LSST, will provide a wealth of data and opportunity for such tests, dubbed ``novel probes''. The phenomenology ranges from supermassive black holes to galaxy dynamics and galaxy cluster profiles. The physics ranges from modified gravity to dark matter interactions and the detection of string theory-motivated scalar fields. Starting with the Snowmass report of 2013, a number of community reports have highlighted this opportunity~\cite{2013arXiv1309.5389J,2016arXiv160407626D}. 
To take novel probes research from scattered individual efforts to a more systematic and productive program requires three elements: a methodical data analysis plan that uses rigorous techniques for mitigating systematic uncertainties comparable to those for the large-scale cosmology analysis; adaptation of existing simulations to construct mock data to test such analyses; support for personnel who can translate  theoretical ideas into innovative small scale tests. The first element can be realized within survey projects, while the latter two require new effort at the theory-simulation-observation interfaces that can benefit multiple projects simultaneously.  

\item Preparing for next-generation experiments. 
What we learn in the next ten years from DESI and LSST will inspire the next generation of Stage V experiments.
If LSST and DESI do find evidence for physics beyond a cosmological constant, we will need to pursue a new observational program that leverages all modes so far unexplored to resolve those exciting results. If the Stage IV experiments find consistency with a cosmological constant, we will have to test this finding at even higher level of precision and fully test the assumptions of General Relativity in the cosmological model. 
The simplicity of the background cosmology in the scenario of a cosmological constant would also enable Stage~V experiments to offer sharper tests of physics such as the nature of dark matter, neutrinos, inflation, and dark radiation candidates.  Well-understood advantages and limitations of the most advanced technologies used in Stage~IV gives us foresight to begin development today toward the technologies that will be needed for a Stage~V program.
In parallel with advancing hardware technologies, we need to advance our theoretical understanding of the origin of the cosmological constant and develop new probes that can find compelling alternative explanations that can be explored with better measurements. 
As a whole, we need to advance our technologies to enable the most comprehensive Stage~V observations and enhance our theory, modeling, and simulation capabilities to inform the design of the best possible programs.

\end{itemize}

These enhancements to LSST/DESI science and the potential for larger-scale, dedicated dark energy programs motivate our Small Projects portfolio.  We describe the potential to  integrate complementary observations into DESI/LSST core projects in Section~\ref{additionaldata}.  We describe new technology developments that will enable Stage~V dark energy experiments in Section~\ref{newtech}. Finally, we present a plan for advancing theory and simulations in Section~\ref{sec:theory}.  
In total, these three complementary efforts fit well within a cohesive Small Projects portfolio dedicated to dark energy science.  
Our Section~\ref{additionaldata} is perhaps most open to new ideas since it encompasses a large number of very different approaches. Sections~\ref{newtech} and \ref{sec:theory} are somewhat more prescriptive as they reflect a stronger consensus on priorities in the two workshops.


\section{Enhancing LSST and DESI with Complementary Data}
\label{additionaldata}


The unifying idea in this section is that there are multiple 
complementary datasets that can enhance LSST and/or DESI by reducing some of the limiting systematic uncertainties in their dark energy analyses or by enabling certain analyses that would not otherwise happen with LSST or DESI alone. 
While LSST and DESI can carry out their baseline dark energy analysis without these external datasets, having the external data indisputably enhances and makes more robust these surveys by providing more room in the systematic error budget for other unanticipated issues that may arise and by enhancing the survey capabilities in other ways.

DOE-supported researchers are focused on the high-level science goals of the core DOE programs and are not typically supported to supplement those studies with additional data samples. Increasing support and flexibility for PIs to bring 
postdoctoral researchers into their groups specifically to leverage other (non-DOE) facilities has the potential to greatly enhance LSST/DESI. The projects are divided into those that involve supplementary datasets that would need to be acquired through some means (Subsection~\ref{complementary}) and those that will enable better cross-survey coordination or some other synergy with already-planned surveys (Subsection~\ref{bridge}).  The typical scale of the projects is $<$\$1M.

\subsection{Complementary Measurement Efforts}\label{complementary}

All projects in the category of Complementary Measurement Efforts will advance the P5 science drivers by integrating new data that add value to the LSST/DESI samples.  These new data will not provide independent dark energy results but will rather allow new or refined measurements with these Stage IV experiments.  Programs should be given preference for improving the calibration of Stage IV data, increasing the dimensionality of Stage IV data for its measurements in weak lensing, galaxy clusters, galaxy clustering, or type Ia SNe, or enabling new cosmological probes with Stage IV data that would not otherwise be possible. Below are several compelling examples of programs involving LSST or DESI along with some other (non-survey) data:

\begin{itemize}
\item {\bf Photometric calibration:} The Type Ia supernova technique requires that both high- and low-redshift SNe be on the same flux system. Only point-like sources can precisely follow the same telescope optical path as the supernovae. Therefore standard stars are needed whose flux calibration as a function of wavelength is on a physical system to an accuracy much better than 1\%. Current photometric systems have high internal consistency, but their wavelength-dependent calibration is only accurate to a few percent. A program to establish the “absolute-relative” flux calibration by referencing standard stars to NIST-traceable light sources is needed. While LSST will attempt to provide an internally-consistent photometry system, there is broad agreement that it will not be able to establish the ``absolute-relative'' flux calibration system.  In particular, to calibrate out LSST filter shifts across the LSST field of view or over the 10-year LSST survey,  this calibration needs to be obtained spectrophotometrically. Moreover, to fully exploit the synergy of LSST and WFIRST, this calibration needs to be extended to NIR wavelengths. Despite the technical challenges, the instrumentation and an associated telescope to carry out this program would have a modest cost, perhaps under \$500K.

\item {\bf Peculiar velocity studies:} 
Measurements of peculiar velocities can be used as a probe of structure growth. Measurements of supernova distances can be used to map out the peculiar velocity field (e.g., see proof of concept in \cite{2013A&A...560A..90F}). When peculiar velocities are measured in the same cosmic volume as a galaxy redshift survey, cosmic variance cancels out, making this combination uniquely powerful. At lower redshifts, DESI will be dominated by cosmic variance, so a peculiar velocity program at low redshift can greatly enhance the power of DESI. New SN Ia standardization methods, such as comparing ``twin'' supernovae, can measure distances to $\sim$3\%. Several thousand such supernovae over the DESI footprint (such as those now being found by ATLAS and ZTF) could be obtained in just a couple of years with a modest investment in the operating costs of existing facilities. LSST will become a source of nearby SNe in the southern hemisphere, and those could be paired with the southern portion of the DESI Bright Galaxy Survey or with a new southern spectroscopic redshift survey. The resources for a LSST nearby SN follow-up program could be established from a pilot SN Ia peculiar velocity program in the northern hemisphere, coordinated with DESI. Such peculiar velocity programs will also establish the velocity zeropoint error, which will set the floor on the accuracy of the anchor for the LSST and WFIRST SN Ia Hubble diagrams, and the accuracy of $H_0$ attainable using gravitational wave optical counterparts.

\item {\bf Narrow-band or offset broad-band imaging:} Well-calibrated photometric-redshift estimates are critical for LSST dark energy science. Narrow-band imaging or imaging with offset broad bands can lead to significant improvements in photometric redshift estimation \cite{1538-4357-692-1-L5}, especially the reduction of the outlier rate.  While narrow-band imaging programs covering a substantial fraction of the LSST footprint would be an expensive and time-consuming proposition, it is possible that offset broad-band imaging over a small portion of the LSST footprint could provide some of the benefits at far lower cost.  Support for efforts to design, e.g., a DECam-based, offset broad-band imaging survey that would reduce systematic uncertainty due to photometric redshifts for LSST would be needed in order to do this, along with the costs of the filters themselves.


\item {\bf Ground-based Spectroscopy:} Several ground-based spectroscopic facilities that would complement LSST and DESI are now in the construction or design phase.  The Subaru Prime FocuS Spectrograph \cite{2016SPIE.9908E..1MT} will begin operations in 2019.  Looking further ahead, a fiber-based Wide-Field Optical Spectrograph is being studied for the Thirty Meter Telescope, the Maunakea Spectroscopic Explorer \cite{MSE} is planned as an 11.25m aperture wide field spectroscopic survey telescope, and the Giant Magellan Telescope is envisioned as a facility that will span 320–25000 nm with a collecting area equivalent to a 24.5-meter telescope.  Depending on the goal of the observations, in some cases these ground-based spectroscopic facilities have advantages over DESI such as better sensitivity to high redshift, faint galaxies and potential for targeted observations.  As a result, data from these powerful instruments could establish a spectroscopic sample for training LSST photometric redshifts \cite{2015APh....63...81N}, enable classification and redshifts for Type~Ia supernovae, provide dynamical information for cluster galaxies, enhance gravitational time delay measurements \cite{2016A&ARv..24...11T}, and offer other forms of spectroscopic insight into dark energy science.  While operations would likely exceed the budget for a program in this small projects portfolio, it is possible that the LSST Project could offer access to LSST data in exchange for spectroscopy as an in-kind contribution.  If this takes place in the coming years, then a modest investment from this small projects portfolio for personnel costs would enable the design of observations and analysis, thus ensuring that the data are used as effectively as possible for dark energy science.

\end{itemize}

The above list is intended to provide representative examples (rather than a complete list) of smaller, targeted observational programs that would enhance LSST and DESI. 

\subsection{Bridging Surveys}
\label{bridge}
Here we describe efforts that will pave the way for enabling cross-survey science, as described for example in Refs.~\cite{2017ApJS..233...21R,2015arXiv150107897J}, which outlines scientific synergies for LSST, WFIRST, and Euclid or Ref.~\cite{2017arXiv171009465S}, which focuses on CMB-S4 and LSST. Combining survey data will require preparatory work that would be part of our Small Project portfolio. It is important that this work start as soon as possible to ensure that the survey data and observation plans are developed in ways that enable these analyses.  
As mentioned above, DOE support tends primarily to focus on the high-level science goals of the core DOE programs and researchers are not typically supported to explore strategies for cross-survey synergy especially with non-DOE projects as described here.
Hence in the majority of cases we are advocating for research funds specifically dedicated to personnel support to develop methods for the cross-survey combinations described below.  Typically the goal of the proposed study is to substantially mitigate some key systematic uncertainty so as to provide more margin in the systematic error budget for LSST or DESI:

\begin{itemize}
\item \textbf{WFIRST$+$LSST}: The fact that WFIRST will have spectroscopy as well as imaging at NIR wavelengths leads to a number of interesting cross-survey synergy opportunities.  Taking advantage of these opportunities requires work on optimal strategies for doing so, and significant preparatory development and testing of actual robust pipelines, since any work that coordinates with space-based efforts must be reviewed and certified well in advance for such collaborations to be permitted.  There are a few different aspects to develop:
\begin{itemize}
\item \textbf{Supernova imaging and spectroscopy}: The LSST survey will, over 10 years, provide a few hundred thousand supernovae from the wide survey, and another few tens of thousands from the deep-drilling fields.  WFIRST NIR imaging and spectroscopic follow-up with the integral-field channel for LSST SN discoveries at $z<0.8$, while they are still active, can provide a key sample of $\sim$3000 SNe Ia with optical-through-NIR spectrophotometry and NIR imaging (and host-galaxy redshifts). This would make possible important tests of SN population evolution and dust evolution that are likely in the end to be limiting sources of systematic uncertainty in the SN dark energy measurements.  At the same time, this spectrophotometric SN subset would make it possible to train the photometric classifiers that can be used for the larger samples of LSST's photometry-only SNe.
All the LSST SN cosmology results could thus be made significantly stronger statistically, and with key controls on systematic uncertainties that make all the difference when/if the next generation of DE measurements show similar tensions between measurement techniques as are seen in today's generation of measurements.  (Discovering the $z<0.8$ SNe with LSST will also significantly enhance the statistical reach of the combined LSST-WFIRST effort, since at these lower redshifts the larger LSST imager field-of-view is a more efficient discovery tool.)

\item \textbf{Spectroscopy for photometric redshift training and/or calibration}: If WFIRST can carry out NIR spectroscopy in parallel mode during the high-latitude imaging survey, it should be possible to build up a training sample with up to $10^5$ galaxies in regions of color-space where ground-based spectroscopy is very challenging.  Ideally this would use methods similar to those employed by the C3R2 survey \cite{2015ApJ...813...53M,2017ApJ...841..111M} to ensure optimal target selection that makes the most of the available time to fill in the photometric redshift training sample. 
\end{itemize}
In order to ensure feasibility of these programs, work on their design needs to start well in advance of the launch of WFIRST.
\item \textbf{LSST and DESI $+$ CMB S4}: There are a number of areas of scientific gain from the combination of CMB S4 and LSST \cite{2017arXiv171009465S} or DESI (within the constraints imposed by limited area overlap). These include use of CMB lensing cross-correlation with LSST to provide an external calibration constraint on the combination of shear and photometric redshift uncertainties \cite{2017PhRvD..95l3512S} and to constrain structure growth to higher redshifts, given the shape of the CMB lensing kernel.  The combination of LSST, DESI, and CMB-S4 is even more powerful, given that including redshift-space distortions in the measurement provides a way to distinguish between dark energy and modified gravity \cite{2007PhRvL..99n1302Z}. This work will require cross-correlation measurements to far smaller scales than the DESI BAO measurements, which connects to the proposed work on modeling in Section~\ref{sec:theory}.
\end{itemize}

An additional area for discussion in the community is the topic of joint pixel processing between WFIRST, Euclid, and LSST (c.f.~\cite{2017ApJS..233...21R}).  As argued there, this has the potential to benefit LSST especially regarding the issue of deblending and associated shear and photometric redshift uncertainties.  However, the quantitative cost/benefit analysis for this joint pixel processing is not yet complete.  It is worth revisiting the question of whether the Small Projects portfolio should include joint pixel processing once that cost/benefit analysis has been carried out by the task force appointed by the Tri-Project Group.



\section{New Technology Developments}
\label{newtech}


Much of the recent rapid progress in cosmology can be attributed to new experiments enabled by instrumentation technologies developed over the last decade.  These efforts followed a shared philosophy with the latest P5 report to pursue technology development in a balanced mix that addresses short-term, immediate need and long-term R\&D.  Likewise, the R\&D efforts that made possible the Stage III and Stage IV dark energy experiments were pursued through partnerships between national laboratories, universities, and the private sector.  Examples include CCDs, fiber positioners, and CCD electronics and data acquisition hardware.

Dark energy experiments depend critically on thick, high resistivity CCDs that provide very high quantum efficiency in the near-infrared.  CCD development activities for dark energy science were performed at the Lawrence Berkeley National Laboratory (LBNL) in partnership with DALSA Semiconductor\footnote{https://www.teledynedalsa.com/semi/CCD-fabrication/}.  Over a ten-year period, these deep-depletion CCDs advanced from an early prototype to a device that can be manufactured in bulk with few cosmetic defects and nearly optimal sensitivity across the wavelength range $3,600<\lambda<10,000$ \AA \cite{2006SPIE.6276E..0BH}.  Scientists at Fermi National Accelerator Laboratory (FNAL) developed the system to efficiently test, characterize, and identify devices for the very large DES focal plane \cite{2006SPIE.6269E..3KE}.  These detectors are now the core technology in the DES, BOSS, and eBOSS cosmology experiments.  

DESI is nearing the end of construction and will expand the spectroscopic reach of BOSS and eBOSS by an order of magnitude.  The integration of 5,000 robotic fiber positioners into the focal plane of the 4-meter Mayall Telescope is fundamentally responsible for this advancement to a Stage IV spectroscopic survey.  Extensive testing and development at LBNL, in collaboration with University of Michigan, has shown that these newly-developed fiber positioners have superior performance compared to robotic positioners employed at other facilities \cite{2012SPIE.8450E..38S,2016SPIE.9908E..92S}.  Improvements include faster repositioning, simplified anti-collision schemes, and inherent anti-backlash preload.  In a vein similar to CCD development, R\&D for these DESI fiber positioners was well underway more than ten years before the scheduled deployment at the observatory \cite{2008SPIE.7018E..50S}.

LSST will host a 3.2 giga-pixel array, the largest ever employed in any astronomy or cosmology facility.  The projections for LSST cosmological constraints from weak lensing and other cosmological probes are based on strict requirements for image quality at a pixel data rate of roughly 3.5 Gigabytes/second, a significant challenge.  The LSST Camera was designed by a consortium led by the SLAC National Accelerator Laboratory (SLAC) and including groups at Lawrence Livermore Lab, Brookhaven National Laboratory (BNL), Harvard, Paris, University of Pennsylvania, and others.  BNL began development toward the LSST focal plane, including the creation of its production facility, in 2003.  The core technology for the Data Acquisition (DAQ) system is derived from SLAC’s ongoing Detector R\&D program. That program has produced a set of generic, modular building blocks (a ``tool-kit'') used to construct scalable, DAQ systems.  Testing shows that the system can now read all 3.2 giga-pixels in less than two seconds while exceeding requirements on readnoise, cross-talk, quantum efficiency, cosmetic quality, diffusion, and heat production \cite{2016SPIE.9915E..0XO}.  

Motivated by the successful development of technologies for Stage III and Stage IV programs, the 2016 White Paper by the Cosmic Visions Dark Energy group \cite{2016arXiv160407821D} provided a summary of possible technologies that could be developed for the Stage V generation of dark energy experiment.  Suggestions in that work include improvements to silicon CCD performance, scale production of germanium infrared imagers, multi-color CCDs, Microwave Kinetic Inductance Detectors, Ring Resonators, Multi-Conjugate Adaptive Optics systems, and improved fiber positioners.  Following a year of reflection since that White Paper and two community workshops\footnote{http://cvde2017.lbl.gov/; http://kicp-workshops.uchicago.edu/FutureSurveys/}, the community has identified three specific technical improvements that can lead to significant enhancements of LSST or another order of magnitude advancement in spectroscopic survey power.  We describe an experimental program to tackle these three technological hurdles.  We estimate the costs for each of these three R\&D efforts to range from \$1M to \$2M:

\begin{itemize}
\item  {\bf Ground Layer Adaptive Optics}:  Ground-based imaging and spectroscopic surveys suffer degraded resolution due to turbulence in the ground layer and in upper layers of the atmosphere.  By correcting for this turbulence in real time, adaptive optic systems can yield substantially improved angular resolution and significant improvements in the signal-to-noise ratio. For example, an improvement from seeing of 0.7 arcseconds to 0.5 arcseconds will yield a factor of two improvement in signal-to-noise for the faintest objects in any imaging program.

Multi-Conjugate Adaptive Optics systems (MCAO) use multiple natural or laser guide stars with several deformable mirrors. The GeMS instrument at Gemini South \cite{2014SPIE.9148E..0YD} achieves diffraction-limited imaging across a field of view of more than 1 arcminute. However, these fields are too small for cosmological surveys and the required large number of galaxies.  

Ground Layer Adaptive Optics (GLAO) offers an alternative to those systems that try to reach the full diffraction limit of the telescope.  In GLAO, wavefront sensors assess common ground-layer turbulence over a much larger field of view using bright guide stars as a reference.  One such GLAO system has been tested on the University of Hawai'i 2.2-meter telescope on Maunakea, Hawai'i. The ``imaka'' GLAO pathfinder system has been shown to produce images with FWHM of 0.33 arcseconds in the visible and near infrared over a 0.33 degree field of view.  Equally important to the improved resolution is the temporal uniformity of resolution recorded by the imaka system.  The RMS in the FWHM from exposure to exposure is reduced relative to the non-corrected seeing, but further testing needs to be done to fully quantify the effect and dependence on observing site.

The imaka system offers proof of concept to the capabilities of GLAO systems to improve image quality over a large field of view.  LSST, DESI, and future cosmology surveys all push for fields of view an order of magnitude larger than the 0.33 degree diameter system being tested now in Hawai'i.  There is only one vendor working on the technology for large GLAO mirrors\footnote{http://www.adoptica.com}, so DOE support would spur further advancement in the field.  

A development effort between 2020 and 2025 could (1) demonstrate the science gains from such systems, (2) characterize the trade-offs between telescope aperture size, field of view, and GLAO improvements, and (3) further develop the technology for wide-field GLAO suitable for DOE science.  If shown to be feasible, such a system could potentially be employed through an updated secondary mirror at LSST.  Such a deployment, even at the midway point in the survey, would significantly enhance the reach of LSST toward faint sources at high redshift that are needed to optimize lensing measurements.  An implementation of GLAO on the Mayall primary mirror for an extension of DESI or in a future spectroscopic facility would equally benefit from such a system;  studies of diffraction-limited data from the Hubble Space Telescope indicate that improvements to 0.6 arcsecond FWHM will optimize sensitivity to high redshift galaxies in a fiber-fed spectrograph.

\item  {\bf Germanium CCDs}:  Silicon CCDs have matured for optical bands covering $3,600<\lambda<10,000$ \AA, but the effective band-gap around 1 eV limits their effectiveness at redder wavelengths. The primary spectroscopic feature used to determine redshift in galaxy surveys is due to forbidden transitions in singly-ionized oxygen ([OII]).  These [OII] emission lines occur at 3727 \AA\ in the galaxy restframe, causing the signal to appear beyond the 10,000 \AA\ cutoff in a silicon detector for galaxies at redshifts $z>1.6$.  Enormous, relatively unexplored volumes will still be available at these higher redshifts even after DESI is completed.  Many physical models (such as early dark energy) are best explored by contrasting the expansion history and growth rate at these high redshifts with measurements of the same in the local Universe.   Detectors with sensitivity at wavelengths longer than 1 micron will extend galaxy surveys to these higher redshifts. 

While infrared InGaAs and HgCdTe CMOS detectors have been used in ground- and space-based observatories, these detectors are expensive, require substantial cooling, and suffer from low yield in the fabrication process.  An alternative to CMOS detectors has recently been identified through work performed at MIT Lincoln Laboratory.  Germanium CCDs can be processed with the same tools used to build silicon imaging devices, show promise for read noise and sensitivity comparable to that of silicon detectors, and offer a high quantum efficiency to wavelengths as red as 1.4 microns when cooled to 77 K.  This increase in wavelength coverage will allow a spectroscopic identification of [OII] emission lines to $z=2.6$, a factor of two increase in volume over what is accessible in the DESI galaxy sample.

Fabrication of germanium CCDs faces several challenges that need to be addressed before these devices can be integrated onto large focal planes.  Several processes in doping, etching, and film deposition are similar to those in silicon CCD fabrication, and may be compatible with DALSA's capabilities.  However, water solubility and low-temperature limitations result in the need for changes in gate-electrode technologies.  In addition, there is only one wafer vendor in germanium and further investigation is required to ensure that purity requirements can be met at scale production on large wafers.  Finally, germanium is higher density than silicon and requires a full assessment of handling and packaging techniques.  We estimate that five-years of effort are required to develop a manufacturing pipeline that can produce the first packaged germanium CCD's for dark energy surveys.

\item  {\bf Fiber Positioner Systems at 5 mm Pitch}:  
The BOSS and eBOSS surveys have collectively sampled the spectra of more than three million objects, each requiring an individual fiber optic supported by a custom aluminum plate.  These fibers have been manually placed by dedicated technicians on an almost daily basis since 2009. The process of plugging typically takes about 30 minutes for each of the $\sim 3000$ fields that comprise the BOSS and eBOSS cosmology samples.  The massive overhead of human effort inspired the investment toward new robotic fiber positioners for future experiments even before BOSS started observations.

The fiber positioners for DESI consist of 5000 individual robots supported by a 812 mm diameter aspheric focal plane at a 10.4 mm pitch between neighboring units.  Each of these robots is driven through two rotational axes by independent, brushless 4mm diameter DC gearmotors from Namiki\footnote{http://www.namiki.net/sp/dcmotor.html}.  The size of these motors, fasteners, and mounting interfaces places a hard limit of roughly 10 mm pitch between units.  The assembly is well underway, but is complicated by the tight spacing between positioners, need for manually applied glue joints, splicing of fibers, and large number of individual parts (675,194 in total).  

Following DESI and LSST, the potential exists to make major advances in constraining the cosmological model from a massive spectroscopic program.  With spectroscopy of 40,000 galaxies per square degree, a dedicated facility could obtain a clustering sample that contains almost all of the cosmological information to redshifts $z<3.25$. A project of such scope is another order of magnitude increase in capability over DESI and requires a comparable scaling of fibers in the focal plane.  Given the current limitations of optics to roughly 1.2 meter in diameter, new fiber positioner technologies will likely be required to populate the focal plane at a sufficiently high density.  A technology that allows fiber spacing at a 5 mm pitch over a 1.2 meter focal plane would allow simultaneous spectroscopy from 50,000 fibers.

An increase in multiplexing from DESI's 5000 fibers should be developed either as an upgrade of the DESI focal plane (with a spacing of 5 mm or less) or an optical/mechanical solution on other telescopes.  It may be possible to improve the two axis DESI positioners by using smaller motors, press-fit joints instead of glue, alternatives to splicing, and new approaches to handling a one ton focal plane assembly.  On the other hand, the ``Tilting Spine'' technology has been used in optical spectrographs and was selected as the fiber positioner technology for the 4MOST facility at the ESO VISTA telescope \cite{2016SPIE.9911E..1YB}.  The Tilting Spines position the end of the fiber by tilting about a long axis rather than positioning in a plane about a central coordinate.  This technique potentially allows for more close packing capability, but because the end of the fiber moves on an arc, larger tilt angles can lead to light loss and irregular injection.  We encourage a comprehensive study and new design for fiber positioners over the period 2020--2025.  If new designs can achieve a goal of 5 mm pitch (half that of DESI), then the most significant technological hurdle for a stage V spectroscopic program can be retired.

\end{itemize}

Following the P5 recommendations and precedent from the last decade, each of these programs should be pursued through partnerships between universities, national laboratories, and the private sector. For example, adaptive optics technologies are already being developed and tested in universities with well-recognized graduate programs, but have not yet been developed for dark energy experiments.  Fabrication of detectors and fiber positioners for DESI is at peak activity, leading to deep expertise in all three of these communities.  This expertise can soon be leveraged and re-directed toward detector and fiber positioner R\&D.  If successful, these efforts will lead to new, cost-effective capabilities in the mid-2020's, in time to be utilized in an upgraded LSST or DESI program or in Stage V dark energy experiments that will be identified in the next P5 process.

\section{Theoretical and Simulation Advances}\label{sec:theory}

Simulations play a central role in modern cosmology in many ways -- they allow us to build virtual universes in which we can explore new physics beyond $\Lambda$CDM, investigate systematic uncertainties and associated mitigation strategies, and develop and test new probes. Predictions from simulations are crucial ingredients in obtaining cosmological constraints from the observations. Developing efficient simulation codes and tools to convert the simulations into well-validated, synthetic skies entails major efforts that require input from theorists and observers alike. The problems posed by the complexity of such efforts can only be addressed through a strong lab-university partnership. For example, SLAC and Argonne National Laboratory in collaboration with university partners and other Labs have built strong simulation and modeling programs over the years. These efforts contribute to all DOE-supported  dark energy experiments.

The importance of cosmological simulations and computing was highlighted in the P5 report under Enabling R\&D and Computing and in the more general Recommendation 29 that addresses computing and simulation needs of the HEP community. The Decadal Survey 2010 also made explicit recommendations for strong support of such a program. Given the importance and the exciting scientific opportunities opened up by comprehensive modeling and simulation efforts, the third component of our Small Projects portfolio describes four concrete areas with potentially high impact. 

The first area concerns the exploration of the deeply nonlinear regime of galaxy clustering. Such an effort would open up new opportunities to extract cosmological information from available observations that otherwise will be lost. Carefully modeling the galaxy-halo connection on these scales is the key to unlock this information. Next, we will describe an effort that would allow us to study and develop novel probes that go beyond the current $\Lambda$CDM paradigm. A third area of potentially great impact would be a coherent effort in building a multi-wavelength virtual observatory. Such an effort would allow us to take full advantage of the ``Bridging Surveys and Wavelengths'' program, described in Section~\ref{bridge}.
Finally, the program could support building a new infrastructure to share simulations and tools among different surveys and therefore lower the overall computational burden for the Cosmic Frontier.
We describe in more detail below such a simulation and theory program. The first three efforts listed could be carried out at a cost of \$1M-\$2M each, while the infrastructure program is expected to lie in the \$2M-\$3M range.

\begin{itemize}

\item {\bf Unlocking Small Scales:}
The scientific potential for extracting cosmological information from small scales has been appreciated for a long time: {\em``The nonlinear domain appears to be a gold mine of cosmological
information, but one whose riches may prove extremely
difficult to extract''}~\cite{1997PhRvL..79.3806T}. Recently, Krause and Eifler \cite{2017MNRAS.470.2100K} carried out a realistic FoM forecast in the present-day context. They showed that by extending a cross-correlation analysis of LSST galaxy clustering and lensing into the highly nonlinear regime, FoM gains by a factor of 2-4 can be realized. During the 20 years between these two publications, the resolution in simulations has increased by orders of magnitude, the statistics (via larger volume simulations) has been improved, and the available observational data sets for validating our modeling approaches on small scales have been vastly enhanced. The time is ripe to therefore attempt to gain access to the cosmological information on small scales.
We stress that this effort only requires more sophisticated modeling and simulation of the standard set of observations already planned by Stage IV dark energy missions such as LSST, DESI, and WFIRST. Therefore, with only a small investment in simulation and modeling efforts, an exciting opportunity exists to enhance the scientific potential of upcoming surveys.

Although the labor involved in extracting cosmological constraints from the nonlinear regime is substantial, a future roadmap can be constructed by melding recent work on detailed modeling of the galaxy-halo connection (see, e.g., Refs. \cite{2015arXiv150703605B}, \cite{2017MNRAS.466.2718C}, \cite{2017MNRAS.471...12B}, \cite{2017arXiv170505373M}) with a large suite of cosmological simulations, both in the gravity-only and hydrodynamics arenas. The roadmap has three components that need to be addressed to bring such an effort to maturity. First, it will be essential to exploit leadership-class computing facilities for the simulation and modeling approaches, the latter of which needs a dedicated program. Second, a major calibration and validation effort will be required. Such an effort includes: (i) calibrating the galaxy population models with enough flexibility to fit a large compilation of observational data; (ii) validating the models against traditional semi-analytic models and hydrodynamical simulations, which will be treated as mock datasets whose underlying cosmology one attempts to recover; (iii) iterating on this procedure using different datasets, refining the model as needed by a range of cosmological data vectors. The third component of such an effort involves using the tools and methodology to forecast the constraining power of nonlinear cosmological observables, to identify cosmological data vectors that can be robustly predicted with the fewest nuisance parameters, and to build likelihood emulators that will generate cosmological constraints. 

\item {\bf Going Beyond $w$CDM:} In Section~\ref{sec:mot}, we motivated the enhancement of a novel probes program to pursue physics beyond $w$CDM. The first part of the enhancement relates to data analysis, here we focus on the second and third: generation of mock data to test analysis codes, and connecting new ideas to observational tests. There is both an opportunity and a challenge in the different  elements of new physics that can operate at small scales. For example, dark matter interactions and gravity can both impact the profiles and dynamics of galaxy clusters. Therefore, in carrying out gravity tests, we must be careful to have robust models of dark matter that extend beyond the simplest CDM candidate – that is the challenge. The opportunity is that by disentangling the precise range of scales and phenomenology we can jointly test for 
dark matter interactions and gravity.

The new physics must be modeled either through full simulations or approximate treatments. A significant opportunity lies in adapting dark energy simulations by approximating the effects of new physics via analytical models. This effort would not require expensive new computations, but rather expertise in topics such as screening effects in modified gravity or 
dark matter interactions coupled with numerical efforts. A related opportunity lies in connecting new ideas to observational tests: a process that currently can take several years owing to the different communities involved. With LSST and DESI offering a new generation of tests, a far more rapid translation of theory to tests can be achieved by providing targeted resources. These could fund collaborative efforts between theorists and observationally oriented cosmologists, with simulations serving as testbeds for new ideas.

\item {\bf A Multi-wavelength Virtual Observatory:} As outlined in Section~\ref{bridge}, cross-correlations of different data sets hold a wealth of information and constraining power. In order to take full advantage of this opportunity, a comprehensive effort in analysis, modeling, and simulations has to accompany the observational campaigns. Such an effort would provide ``same sky mocks'' for the different wavelengths covered by the surveys with proper correlations between all of the signals that will be measured. On the analysis front, an approach has to be developed that utilizes nuisance parameters, bias models, covariances, and systematic models jointly between signals. We will need to investigate the level of abstraction at which we can deal with foreground separation, time-stream filtering, photometric calibration, etc. For this task, simulations will be an essential tool. The ultimate goal is to facilitate joint analysis of simulated data for validation, calibration, and correlation across multiple probes.

Creating a multi-wavelength virtual observatory across all observables, including CMB (projected lensing maps, thermal and kinetic SZ), direct broadband emission across all wavelengths, from far infrared via optical to X-ray emission, and specialized observables such as Lyman-$\alpha$ forest, unresolved 21-cm, damped Lyman-$\alpha$ systems, etc.  is a challenging task. To take full advantage of up-coming observations across these wavelengths, gravity-only simulations at very high mass resolution and hydrodynamics simulations including feedback effects and baryon physics have to be carried out. The gravity-only simulations would be the bedrock for, e.g., the optical sky catalogs, CMB lensing measurements, and also allow the modeling of foregrounds relevant for the CMB surveys, such as the cosmic infrared background. The hydrodynamics simulations will be important for, e.g., modeling the Sunyaev-Zel'dovich effect or cluster cosmology investigations, Lyman-$\alpha$ observables and other small-scale physics. In addition to the simulation efforts, building tools to create synthetic skies are essential. Some of these tools already exist but given the quality of up-coming observations, will have to be sharpened. The validation of these catalogs -- and therefore ensuring their high fidelity -- would be another major task. Some of this would be done at the project level, but close collaborations between the surveys would be essential to fully realize the potential of the multi-wavelength mock catalogs.

\item {\bf Enabling Community Science:} Many of the tools and simulations that the dark energy community is developing can be applied across different surveys. Large-scale simulations require efficient codes and substantial supercomputing resources. Given the limited resources currently available and given the costs of generating these simulations, it is most natural to develop an infrastructure that allows for easy access to such simulations. The establishment of an access point where tools and simulations can be easily shared would be extremely beneficial. Such an effort will require investment in people to build out an infrastructure that is sustainable in the long-term. A recently formed task force is investigating common use of simulations and tools for LSST, WFIRST, Euclid, and DESI and has identified the following important topics for consideration: common infrastructure to share simulation products and tools, a base of numerical simulations to generate synthetic sky maps, tools to generate synthetic sky maps, large-scale simulation campaigns covering different cosmologies and enabling covariance estimates, investigation of systematic effects via simulations (including baryonic effects), and advanced statistical methods. While all of these topics are of great importance, such a program would go beyond a Small Projects Portfolio. We strongly believe that the first item, common infrastructure to share simulations and tools, would have the most immediate impact and focus on this in the following.

An infrastructure for sharing simulations and tools can take on different levels of sophistication, depending on the available resources. The most straightforward implementation would only allow the community to browse the data products and to download the relevant data to their local computing resources. A more comprehensive approach would allow the community to also access computational resources where the data resides to carry out first level analysis. The results, if of general interest, could then be added to the available data for other users. In addition, generic tools could be made available for the analysis of the available data. Finally, the infrastructure could be completed by enabling the community to contribute tools and datasets. Small-scale efforts that could seed such an infrastructure already exist. For example, DESCQA, described in Ref.~\cite{2017arXiv170909665M} provides an environment to contribute to the validation of synthetic sky catalogs, or open source codes for analysis, such as Halotools~\cite{2017AJ....154..190H}, can be used for analyzing available simulations. Building upon smaller efforts like these could lead to a powerful, community driven infrastructure motivated above.

An intriguing opportunity is provided via a possible close collaboration with ASCR in this area. While cosmology could function as a pathfinder project, if built well, the infrastructure could be used and adopted by other offices in the future.

\end{itemize}

The simulation and theory program outlined above covers a range of different aspects, from specific modeling and simulation challenges of small scales, to exploration of physics beyond the Standard Model of Cosmology, to infrastructure development that will reach across different surveys. This ambitious program relies on the establishment of strong university-lab partnerships. If successful, it will enhance upcoming DOE flagship missions in cosmology and at the same time provide the tools to plan and shape next-generation dark energy missions. It also provides exciting opportunities to work across DOE program offices.





\section{Conclusions and Outlook}\label{sec:conclusions}

Following two community workshops in 2016 and 2017, this document outlines the broad consensus on small-scale efforts that can improve the scientific output of DESI and LSST, the two DOE flagship dark energy experiments that will take place in the next decade. In addition to this White Paper the community is also preparing two other White Papers outlining roadmaps for  longer-term development of the large-scale survey science within the DOE HEP: the southern spectroscopic survey roadmap and the 21-cm cosmology roadmap. Both documents are in preparation and will be released in Spring 2018.

Our proposal is informed by significant advances in understanding the dominant sources of systematic errors and advances in the methods of data analysis in the time since these experiments were conceived. While the experiments will deliver their designed science, small inputs of additional funding will make the results significantly more robust. Additionally, new avenues of analysis and data combinations will enable new insight into the nature of dark energy and dark matter. Despite a challenging budgetary environment, the dark energy research is marching forward with vigor. We argue for a balanced but similarly vigorous approach.  The proposed portfolio combines guaranteed science exploring large, theoretically well-understood scales with more speculative approaches that extract information from smaller scales where the statistical and systematic errors from measurements are minuscule compared to theory uncertainties.

We have organized many ideas into three overarching sets: observations that leverage existing facilities, experimental R\&D that will inform design of the next generation of experiments, and a broad program of theory and simulation development. In Appendix \ref{sec:matrix} we have sorted these ideas into a readiness/cost matrix. 

\newpage

\appendix

\section{Project Matrix}\label{sec:matrix}

In the following table, we provide a summary for the possible start dates and rough cost estimates for the different components of our Small Projects Portfolio.

\begin{table}[ht]
\begin{tabular}{|c|l|l|}
\hline
\multirow{2}{*}{Readiness} & \multicolumn{2}{c|}{Total Cost}\\

  &  \multicolumn{1}{c}{<\$1M} & \multicolumn{1}{c|}{\$1M - \$3M} \\
\hline 
<2020 & 
\begin{minipage}{6.7cm}
\ \\
\textit{Extending DESI/LSST}$^\star$:\\
- Photometric calibration instrumentation \\
- Narrow-band or offset broad-band imaging \\
- WFIRST $+$ LSST synergies \\
\end{minipage}
&
\begin{minipage}{6.7cm}
\ \\
\textit{Theoretical and Simulation Advances}:\\
- Modeling \& simulations for small scale clustering \\
- Modeling \& simulations beyond $\Lambda$CDM \\
- Multiwavelength Virtual Observatory \\
- Enabling Community Science \\
\end{minipage} \\
 \hline
2020-23 &  
\begin{minipage}{6.7cm}
\ \\
\textit{Extending DESI/LSST}$^\star$:\\
- Personnel costs for ground-based spectroscopy\\
- Peculiar velocity studies\\
- LSST and DESI $+$ CMB S4 synergies\\
\end{minipage}
& 
\begin{minipage}{6.7cm}
\ \\
\textit{New Technology Developments}:\\
- Ground layer adaptive optics over 10 deg$^2$ field of view \\
- Germanium CCDs manufactured at scale \\
- Fiber Positioner Systems at 5 mm pitch  \\
\end{minipage}\\
\hline
\end{tabular}
\end{table}
$\star$ Less prescriptive category with more scope for new options and ideas.

\bibliographystyle{alpha}
\bibliography{sample}

\end{document}